\documentclass[nofootinbib,twocolumn,aps,pre,preprintnumbers,floatfix,amsmath,amssymb]{revtex4-2}
\usepackage{graphicx,color}
\usepackage{caption}
\usepackage{subcaption}
\usepackage{amsmath}
\usepackage{epstopdf}
\usepackage{soul}
\usepackage{tabularx}
\usepackage{natbib}
\newcommand{\D}{\mathrm{d}}
\newcommand{\e}{\mathrm{e}}

\newcommand{\be}{\begin{equation}}
\newcommand{\ee}{\end{equation}}
\newcommand{\bea}{\begin{eqnarray}}
\newcommand{\eea}{\end{eqnarray}}
\newcommand{\ba} {\begin{align} }
\newcommand{\ea} {\end{align} }

\newcommand{\kbt}{k_{\mathrm{B}}T}

\newcommand {\Na} {\text{Na}}
\newcommand {\K} {\text{K}}
\newcommand {\Cl} {\text{Cl}}

\newcommand{\ct}{\Theta}

\newcommand{\intra}{\text{i}}
\newcommand{\gt}{g_\text{tot}}
\newcommand{\tu}{\tilde{u}}

\begin{document}

\title{Theory of cell size regulation during migration in adhered cells}

\author{Jonathan E. Ron$^{1,2^{*}}$ and Ram M. Adar$^{1^{\dagger}}$}
\affiliation{$^1$ Department of Physics,Technion – Israel Institute of Technology, Haifa 32000, Israel}
\affiliation{$^2$ Department of chemical and biological physics,Weizmann Institute of Science, 234 Herzl st., Rehovot 7610001, Israel}
\begingroup
\renewcommand\thefootnote{}\footnotetext{
$^{*}$ \text{jonathan.ron@weizmann.ac.il} \\
$^{\dagger}$ \text{radar@technion.ac.il}
}
\endgroup
\begin{abstract}
Cell migration is closely linked to cell shape, yet cell size is often assumed to remain constant. This assumption is challenged by recent experiments showing that cells undergo volume loss during spreading and swelling upon activation, with migration velocity correlated to cell size. In this Letter, we present a minimal theoretical framework for cellular size regulation and its influence on migration velocity. We connect cell size to membrane potential and active, actin-driven forces. Spatial inhomogeneities in these forces establish cell polarization and drive migration. Crucially, inhomogeneity is easier to establish over larger sizes, giving rise to a critical contact area, above which migration is possible. Our theory captures the coupled dynamics of cell volume, surface area, and motility and explains recent experiments on neutrophils.
\end{abstract}

\maketitle

\section{Introduction}
 Cellular migration underlies fundamental biological processes such as development, immune response, and metastasis~\cite{Friedl2009,friedl2011cancer,nourshargh2014leukocyte}. From a biophysical perspective, migration requires both a driving force—typically associated with actin polymerization or contractility—and a well-defined polarization that determines the direction of motion~\cite{Hakim2017}.

The role of cell shape changes during migration is well established: asymmetry can facilitate polarization and support persistent motion~\cite{Ron2020,ron2024emergent,liu2024shape,driscoll2012cell,Blanch2013,bodor2020cell,lavi2020motility,sadhu2023minimal}. In contrast, the cell size is often assumed to remain constant \cite{Callan-Jones2013}. However, this assumption breaks down even on short timescales, as cells undergo significant water exchange within seconds and ion transport within minutes~\cite{Cadart2019}. Recent studies have shown that cells lose volume during spreading~\cite{Xie2018,Guo2017,Venkova2022}, a phenomenon attributed to mechanosensitive ion channels that open in response to increased surface tension~\cite{Xie2018,Venkova2022,AdarPNAS,AdarBJ}. Adhered neutrophils have also been observed to swell~\cite{Delpire2018,Ni2024,nagy2024neutrophils} upon chemoattractant exposure via ion channel activation, with migration velocity increasing during the swelling phase~\cite{nagy2024neutrophils}.

In this Letter, we elucidate the coupled dynamics of cell volume and contact area, and their impact on migration over an adhesive substrate. First, we describe volume regulation using the pump-leak model for water and ion transport. Second, we relate spreading dynamics to the interplay between actin-driven protrusive forces and membrane surface tension. Together, volume and spreading determine the contact area of the cell. Finally, we demonstrate that contact area directly influences migration: polarization arises from gradients in actin-regulatory components, which are more readily established across larger spatial domains. This leads to a critical contact area above which persistent migration becomes possible.

Our theory offers a coherent framework for the interdependent roles of electrochemical regulation, mechanical shape control, and motility, and quantitatively captures recent experimental findings on two-dimensional neutrophil migration following activation.


\section{The model}
\label{sec1}
\begin{figure}[ht]
\centering
\includegraphics[width=0.98\columnwidth]{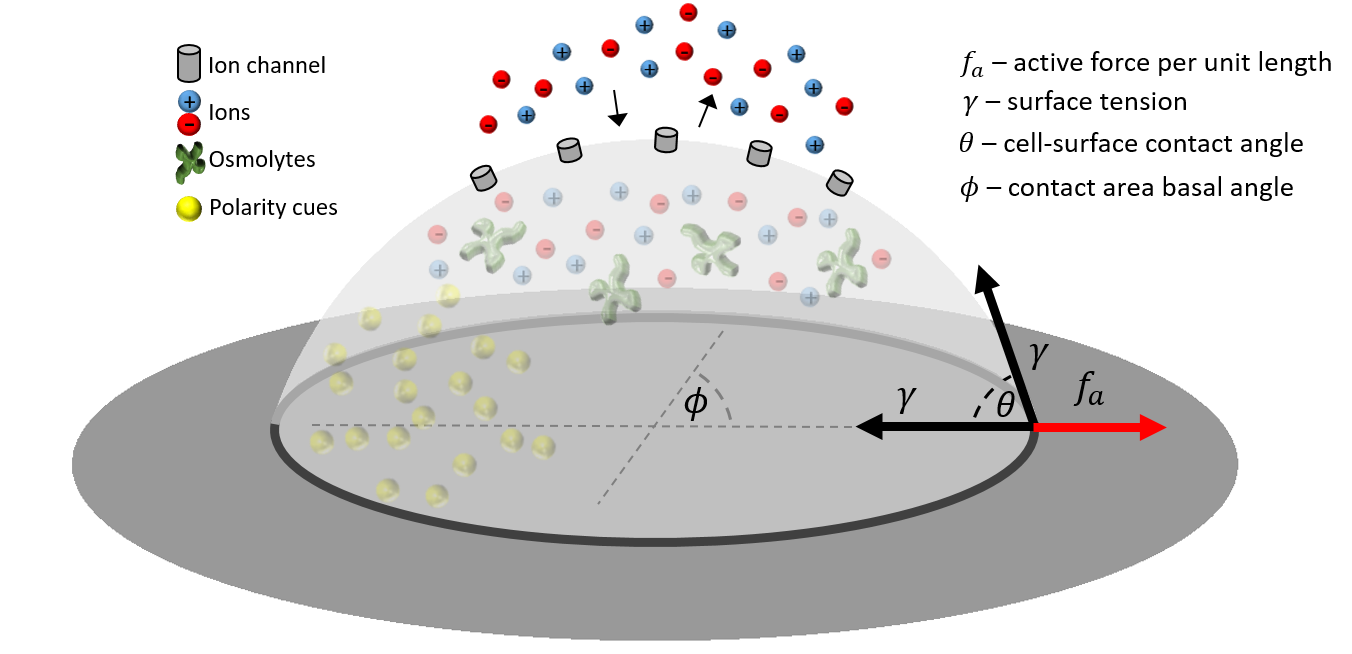}
\caption{Model illustration. An adhered cell is represented by a spherical cap. At each point along the contact ring, an active force per unit length $f_a$ acts outwardly, and is opposed by the cellular surface tension $\gamma$. The angle $\theta$ is the cell-substrate contact angle and the angle $\phi$ is the azimuthal angle along the contact ring. Red and blue mark ions that exchange with the extracellular environment. Green marks impermeant molecules. Yellow marks actin regulatory polarity cues.}
\label{fig1}
\end{figure}

We consider a simplified model of a cell adhering to an underlying substrate and adopting the shape of a spherical cap (see Fig.\ref{fig1}). The cell interior is modeled as a solution, comprising of water, ions that exchange with the extracellular environment, and impermeant molecules. 
Among the impermeant molecules are proteins that interact with cortical actin and act as polarity cues (see Fig.~\ref{fig1}).


Next, we apply this model to derive coupled equations for cellular volume, contact angle, and migration velocity.

{\bf Cell volume -} We find the cellular volume $V$ via the pump-leak model~\cite{Tosteson1960,Keener,Kay2017}, which accounts for active ion transport through pumps and passive ``leak'' down the electrochemical potential gradient through ion channels (see Appendix A). The volume is determined by osmotic pressure balance with the extracellular environment and is given by~\cite{AdarPNAS,AdarBJ,AdarXiv2024}
\begin{equation} \label{V(t)}
V(t) = V_d+\frac{V_m-V_d}{1-\e^{\psi(t)}}.
\end{equation}
Here $V_d$ is the dry cellular volume due to the impermeant molecules and intracellular membrane (obtained experimentally at infinite extracellular pressure) and $V_m$ is the minimal possible volume in isotonic conditions, obtained when the cell contains only the impermeant molecules and their counterions. The contribution of additional ions is tuned by the dimensionless membrane potential $\psi<0$. Note that for extremely hyperpolarzied cells ($\psi\ll-1$), all the additional ions escape the cell and $V=V_m,$ while extreme depolarization ($\psi\approx0$) will lead to very large volumes and, practically, cell rupture. 

The membrane potential enforces intracellular electrnetruality in the presence of ion transport. Its value is determined by the ionic pump rates and passive conductances, as well as the intracellular and extracellular ionic concentrations.  We consider potential dynamics due to two sources of changes in ionic conductances: (i) mechano-sensitive (stretch-activated) channels that open during spreading due to increased surface tension~\cite{Phillips2009,Xie2018,AdarPNAS,AdarBJ,Venkova2022}, and (ii) channel activation as part of regulatory volume increase (RVI). The latter is relevant to volume adaptation after hypertonic osmotic shocks~\cite{Hoffmann2009,Delpire2018} and upon activation of neutrophils to chemo-attractants~\cite{nagy2024neutrophils}.  For simplicity, we consider instantaneous adaption of the channels.

Within linear response, the potential evolves according to (see Appendix A)
\begin{equation} \label{psi}
\psi(t>0)=\psi_0-\delta_1\frac{\gamma(t)-\gamma_0}{\gamma_0}-\delta_2\e^{-t/\tau_\psi}.
\end{equation}
where $\psi_0$ is the steady-state value after long times, $\delta_1$ accounts for mechanosensitivity, where $\gamma(t)$ is the cellular surface tension and $\gamma_0$ is the steady-state tension when the cell is at rest. $\delta_2$ is the change in potential due to activation, while $\tau_\psi$ is the typical adaptation time of membrane potential, related to the leakage of cellular cations~\cite{AdarXiv2024}.
Throughout this paper we focus on the case of $\delta_1>0$, which is supported by experimental evidence from four recent independent studies~\cite{Guo2017,Xie2018,Venkova2022,nagy2024neutrophils} on different cell types, indicating that membrane potential (and hence, cellular volume) decreases with increasing tension. We also focus on $\delta_2>0$, relevant for regulatory volume increase. Under these conditions, the cell swells over a period comparable with $\tau_\psi$.

{\bf Cell contact area -} We consider an adhered cell with the shape of a spherical cap. Its contact area is set by its volume and contact angle $\theta$. The latter is determined by a force balance equation, written in terms of $\ct=\cos(\theta)$, as in Young's law. We consider a cellular surface tension that drives a force per unit length $\gamma\left(1+\ct\right)$ pointing towards the cell. The first contribution arises from the basal cell area, and the second, from the apical area (see Fig.\ref{fig1}). This is balanced by an active force per unit length $f_a$ that acts in the normal direction away from the cell, driven - for example - by actin polymerization.

At long times, we obtain a contact angle of $\ct=f_a/\gamma-1$ (Young's law). At shorter times that are of interest to us, we assume linear-response dynamics, whereby $\ct$ relaxes towards its steady-state value with a finite relaxation time $\tau_{\theta}$,
\begin{align}
\label{eq4}
\left(1+\tau_{\theta}\frac{\D}{\D t}\right)\ct(t)=\frac{f_a}{\gamma(t)}-1.
\end{align}
The tension also changes with time. For simplicity, we assume that the cell maintains tension homeostasis at long times and experiences a transient viscoelastic contribution of a Maxwell model
\begin{align} \label{eq7}
\left(1+\tau_{\gamma}\frac{\D}{\D t}\right)\gamma(t)=\gamma_0+
\eta\frac{\D }{\D t}\epsilon(t).
\end{align}
Here $\tau_{\gamma}$ is the viscoelastic relaxation time, $\eta$ is the viscosity, and $\epsilon(t)$ is the strain $(A_{tot}(t)-A_0)/A_0$, defined with respect to the total cell area $A_{tot}$ (sum of contact and apical areas) and reference area $A_0$.

The contact area is related to the contact angle and volume by the spherical cap approximation
\begin{equation} \label{sphericalcap}
A=(9\pi)^{1/3}\frac{V^{2/3}\left(1+\ct\right)}{\left(2+\ct\right)^{2/3}\left(1-\ct\right)^{1/3}}.
\end{equation}
The contact area dynamics are similarly related to the volume and contact-angle dynamics via the chain rule,
\begin{equation} \label{dAdt}
\frac{\D A}{\D t}=\frac{\partial A}{\partial V}\bigg|_{\ct}\frac{\D V}{\D t}+\frac{\partial A}{\partial \ct}\bigg|_{_V}\frac{\D \ct}{\D t}. \end{equation}
The first term represents area changes driven by volume changes (swelling), while the second term corresponds to area changes due to variations in the contact angle (spreading).

{\bf Volume-area coupling -} The interplay between volume and area is represented by Eqs. (\ref{V(t)}-\ref{dAdt}). We find that the dynamics are governed by two processes: (i) tension driven volume loss ($\delta_1$ term), and (ii) regulatory volume increase ($\delta_2$ term) driven by the activation of ion transporters. Concurrently, the tension depends on the cellular area, which is itself dependent on volume [Eq.~\eqref{sphericalcap}]. Our description applies for timescales longer than the typical timescale of water permeation ($\sim$ 1 second) and of Cl$^-$ leakage~\cite{AdarXiv2024}, which is assumed here to be shorter than the typical spreading time ($\sim$ 1 minute). In addition, we limit ourselves to timescales shorter than an hour, when further regulatory mechanisms become important, such as protein synthesis.

Note that the interplay between volume and area can, in principle, lead to an instability. We demonstrate (see Appendix B) that such an instability may arise only when $\delta_1$ is sufficiently negative, meaning that cells {\it gain} volume upon spreading, rather than lose volume. In this scenario, a small fluctuation in cell volume increases the area, which, in turn, amplifies the volume due to its dependence on tension. 

\begin{figure}[ht]
\centering
\includegraphics[width=1\columnwidth]{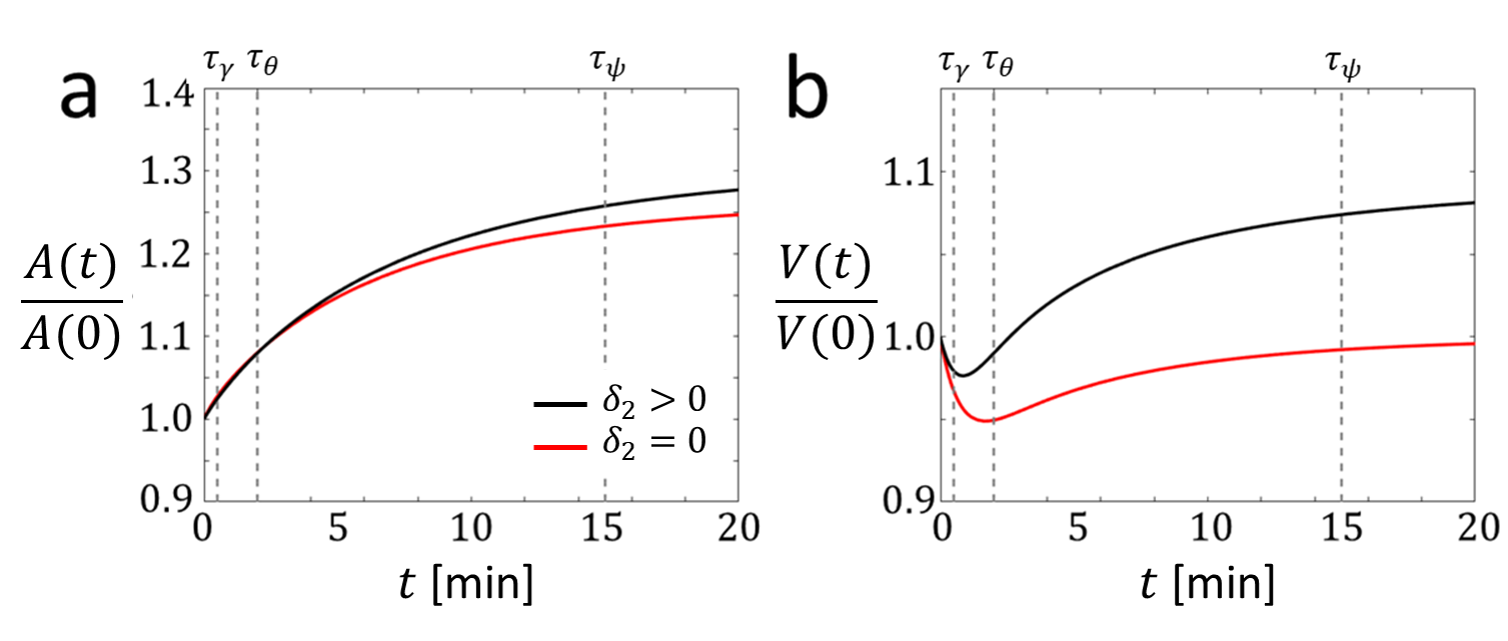}
\caption{Spreading and swelling dynamics, with and without RVI (Black and Red curves respectfully). (a) Normalized contact area.  (b) Normalized volume. Vertical dashed lines indicate the elastic, spreading and potential activation time scales ($\tau_{\gamma}$, $\tau_{\theta}$, and $\tau_{\psi}$ respectfully). Parameters: $\tau_{\gamma},\tau_{\theta},\tau_{\psi}=0.5,2,15$ [min]. $\psi_0,\delta_1,\delta_2=-1.03,4,1.4$ (black) or $0$ (red). $f_a/\gamma_0=1.83$. $\eta A(0)/\tau_{\gamma}\gamma_0=4.8 $. $V_d = 0.3 V(0)$, $V_m=2.5 V_d$. $A(0)=154$ [\text{\textmu m$^2$}].  $V(0)=250$ [\text{\textmu m$^3$}].}
\label{fig2}
\end{figure}

We numerically solve Eqs.~\eqref{V(t)}-\eqref{dAdt} to predict the cellular volume and area dynamics upon activation, as is illustrated in Fig.~\ref{fig2}. We model activation as a sudden increase in the active force per unit length $f_a$ at $t=0$ and consider both $\delta_2=0$ (no RVI) and $\delta_2>0$ (RVI). The dynamics are governed by the three timescales introduced above. The increase in active force leads to an increase in $\ct$ (spreading) and, consequently, increase in area. The timescale of this process is $\sim\tau_{\theta}$. 

During spreading, the tension transiently increases, reaching a maximal value within a time comparable to $\tau_\gamma.$ During this time, the volume decreases due to mechanosensitivity ($\delta_1$ term). Given RVI, the membrane potential increases during a time comparable with $\tau_\psi$ leading to an increase in volume (swelling). Swelling also increases the contact area, while the contact angle remains unaffected.

This establishes the dynamics of the cell volume and area due to the combination of spreading and swelling. Next, we turn our focus to cell migration and show how the migration velocity is largely determined by the contact area.

{\bf Cell migration and polarization -} 
The migration velocity $u$ is determined by the balance between friction and active forces on cellular scale,
\begin{equation}\label{ForceBalance}
2\pi\Gamma u=\int_0^{2\pi}f_a(\phi)\,\cos\phi\,d\phi 
\end{equation}
where $\Gamma$ is the cell-substrate friction per unit length.  As long as the active force distribution $f_a(\phi)$ is homogeneous, there is no net force and $u=0$. Cell migration requires an inhomogeneous force distribution, {\it i.e.}, polarization. In the absence of external fields, polarization can be obtained by a spontaneous symmetry breaking of an internal cellular degree of freedom at some random direction.

In our model, this degree of freedom corresponds to the concentration of proteins that interact with actin along the cell contact ring and regulate actin protrusions in the outward normal direction. We refer to these proteins as ``polarity cues''. We consider generic polarity cues that can represent, for example, membrane curvature regulators such as IRSp53 \cite{scita2008irsp53,mukherjee2022actin}, actin-polymerization inhibitors such as arpin \cite{dang2013inhibitory,UCSP,Ron2020} capping proteins \cite{yang2005mammalian},  or other molecules that activate signaling pathways which promote actin-based protrusions, such as the small GTPase Rac1 \cite{machacek2009coordination,ron2023,bertrand2024clustering}.

Cellular polarization emerges in our theory from a generic feedback mechanism between cellular migration and advection of polarity cues. Similar theories have been successful in capturing key features of cell migration, including velocity and persistence, across different cell types~\cite{UCSP,lavi2020motility,Ron2020,callan2016actin,hawkins2011spontaneous}. The polarity cues diffuse within the cytoplasm with diffusion coefficient $D$ and are advected by the net retrograde cytoplasmic flow, which is assumed to be proportional and opposite to the cell migration velocity~\cite{UCSP,wilson2010myosin,jurado2005slipping}. The interplay between diffusion and advection leads to an asymmetry in the protein concentration along the direction of migration (chosen hereafter as the $x-$direction), which self-consistently drives migration. This interplay is characterized  by the P\'{e}clet number  $\tilde{u}=uR/D$, where $R=\sqrt{A/\pi}$ is the radius of the cell contact area. Based on these arguments, we write the distribution of active forces as
\begin{equation} \label{fa}
f_a\left(\phi\right) = f_0 \left[1+p(\tilde{u}\,\cos\phi)\right]
\end{equation}
where $f_0$ is an intrinsic scale of force per unit length, and $p$ accounts for the anisotropy in the distribution of polarity cues. It is assumed that $p\ll1$, such that its effect on spreading dynamics is negligible [Eq.~\eqref{eq4}].

Next, we insert this expression in the force balance equation [Eq.~\eqref{ForceBalance}] and solve for $\tu$. Due to the many arguments that enter $p$ (such as the protein-membrane interaction, intracellular transport, and force generation description), we phenomenologically expand it in powers of $\tu$, as follows:
\begin{align} \label{treq}
\frac{\Gamma D}{f_0 R}\tu&=\frac{1}{2\pi}\int_0^{2\pi}p\left(\tilde{u}\,\cos\phi\right)\,\cos\phi\,d\phi\nonumber\\
&\approx p_1\tu+p_3\tu^3+p_5\tu^5,
\end{align}
where $p_n$ are polynomial coefficients and where even powers have vanished due to the angular integration. The cell size enters Eq.~\eqref{treq} through the radius $R$ and consequently affects the migration velocity. Its effect is two-fold: (i) It enters the prefactor on the left-hand side and acts in similar to an inverse temperature near a phase transition, and (ii) it defines the typical magnitude of migration speed according to $u=\tu D/R$.

\begin{figure}[ht]
\centering
\includegraphics[width=1\columnwidth]{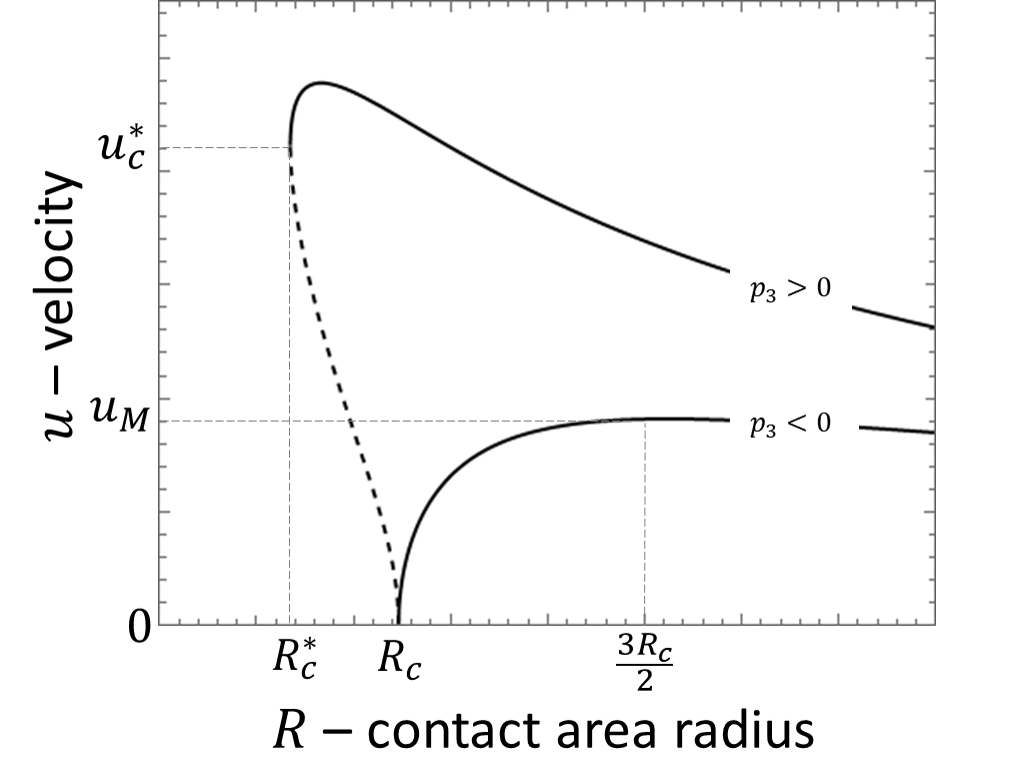}
\caption{Migration velocity profile as a function of contact area radius [Eq.~\eqref{ueq}]. Solid curves indicate the stable solution branches. Dashed curve indicates the unstable solution branch for $p_3>0$. Parameters: $\Gamma D/f_0=1.12$ [\textmu m],$p_3=\pm 1$.}
\label{fig3}
\end{figure}

We analyze the possible values of $\tu$ as a function of $R$ and the coefficients $p_n$. By the choice of the $x-$ axis, we assume that $p_1>0$. We also assume that $p_5<0$, allowing for the truncation of the expansion. The velocity behavior largely depends on the sign of $p_3$, as is illustrated in Fig.~\ref{fig3}. 

For $p_3<0$, we neglect $p_5$ and Eq.~\eqref{treq} reduces to a quadratic equation for $\tu$. We solve it (see Appendix C) and find that $\tu=0$ for small radii and smoothly increases $\tu\ge0$ beyond the critical radius $R_c=\Gamma D/f_0 p_1.$ This smooth transition corresponds to a supercritical bifurcation when velocity dynamics are accounted (Fig.~\ref{fig3} and Appendix C). 
For $p_3>0$, Eq.~\eqref{treq} reduces to a quadratic equation for $\tu^2$. Its solution is $\tu=0$ for $R<R_c^\ast,$ with $R_c^*=R_c/\left(1+p_3^2/4p_1|p_5|\right)<R_c,$ above which there is a sudden jump to $\tu=u_c^\ast R/D>0.$ This discontinuous transition corresponds to a subcritical bifurcation (Fig.~\ref{fig3} and Appendix C).
In both cases, $\tu$ increases with $R.$ However, as $u=\tu D/R,$ the actual velocity exhibits a maximum and decreases with $R$ for large radii, as is shown in Fig.~\ref{fig3}. Explicitly, the velocity is given by
\begin{equation}\label{ueq}
u = \begin{cases}
          \frac{3^{3/2}}{2}u_M \frac{R_c}{R}\sqrt{1-\frac{R_c}{R}}, & p_3<0,\,R>R_c \\
           u_c^*\frac{R_c^*}{R}\sqrt{1+\sqrt{\frac{R_c}{R_c-R_c^*}\left(1-\frac{R_c^*}{R}\right)}} , & p_3>0,\,R>R_c^\ast, \\
\end{cases}
\end{equation} 
and vanishes for lower values of the radius. In the case $p_3<0$, $u_M$ is the maximal velocity, obtained for $R=3R_c/2$. For $p_3>0,$ the velocity suddenly jumps to $u=u_c^\ast$ for $R=R_c^\ast.$ There are two branches: a stable branch that increases with $R$ until a maximum for $R=R_c$ and an unstable branch that decreases with $R$ and vanishes for $R=R_c$ (see Fig.~\ref{fig3}). For the derivation of Eq.~\eqref{ueq} and the dependence of $u_M$, and $u_c^\ast$ on the coefficients $p_n,$ see Appendix C.

Our results have a straight-forward explanation for the size-dependence of the migration velocity: migration requires polarization, achieved within our theory via a gradient of polarity cue concentration along the contact ring. Gradients are easier to establish over larger distances inside the cell, leading to an increase in velocity as the contact area increases. A similar result was obtained in a 1D model of cellular migration~\cite{UCSP,Ron2020}. As polarization is linked within our theory to $\tu=uR/D,$ once maximal polarization is achieved, the velocity eventually decreases with cell area. This happens at relatively large radii ($R=3R_c/2$ in the continuous case), which are not necessarily achievable experimentally. We restrict our discussion hereafter close to the bifurcation, where the velocity increases with radius.
Finally, we highlight that even cells with no net migration velocity $u=0$ are expected to diffuse randomly. Our theory captures this effect by considering anisotropic fluctuations in the distribution of the polarity cues, which modify Eq.~\eqref{fa}, according to $f_a(\phi)=f_0\left[1+p(\tu\cos\phi)+p_{\xi}(\phi)\right]$, where $p_{\xi}(\phi)$ is a random variable that satisfies $\langle p_{\xi}(\phi)\rangle=0$ and $\langle p_{\xi}(\phi)\,p_{\xi}(\phi')\rangle\sim\delta\left(\phi-\phi'\right)$. This provides a correction to the velocity $u_\xi$ with $\langle u_\xi\rangle=0$ and $\langle u_\xi^2\rangle\sim\left(f_0/\Gamma\right)^2$. Interestingly, this result relates the critical area for migration $A_c=\pi R_c^2$ with the mean-squared velocity during diffusion $A_c\sim 1/\langle u_\xi^2\rangle$. This relation can be tested experimentally.

\section{Comparison with experiments} \label{sec4}
\begin{figure*}[ht]
\centering
\includegraphics[width=1\textwidth]{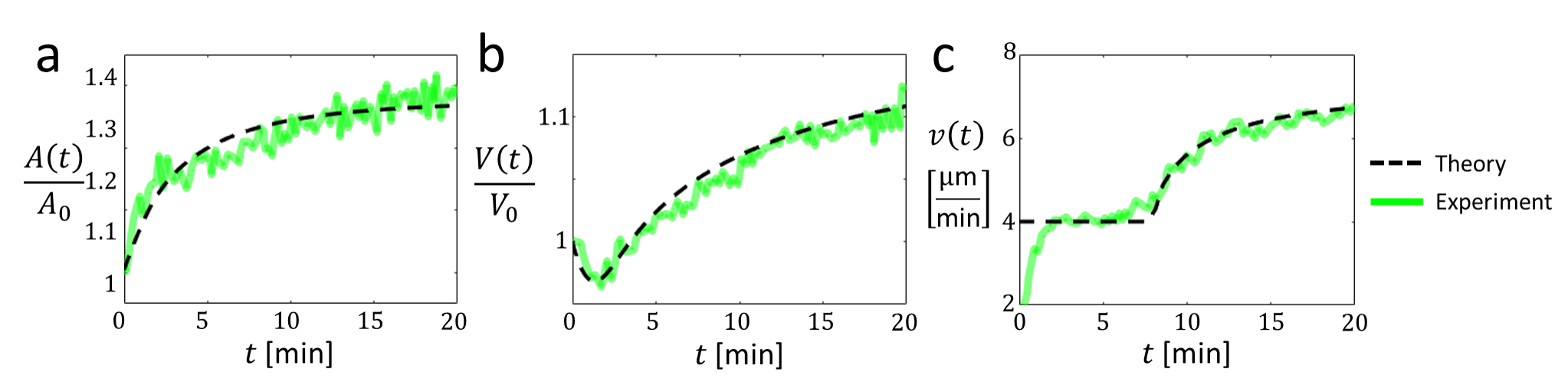}
\caption{Comparison of theoretical predictions (black dashed curves) with experimental data by Nagy et al.~\cite{nagy2024neutrophils} (green curves), for neutrophil area, volume and velocity upon exposure to chemoattractants at $t=0$. a) Normalized area. b)  Normalized volume. c) Cell velocity. Parameters: $\tau_{\gamma},\tau_{\theta},\tau_{\psi}=0.85,0.85,12$ [min]. $\psi_{0},\delta_1,\delta_2=-1.53,0.75,0.08$. $f_0/\gamma_0=1.47,1.96$ (pre/post activation). $\eta/\gamma_0 = 0.54\tau_{\gamma}/A_0$. $\Gamma D/f_0=0.34,0.26$ [\textmu m] (pre/post activation). $p_1,p_3=0.033,10^{-6}$. $V_d=0.3 V_0$, $V_m=2.5*V_d$. Initial conditions: $A_0=154$ [\text{\textmu m$^2$}], $V_0=250$ [\text{\textmu m$^3$}], $\Theta_0=0.44$.}
\label{fig4}
\end{figure*}

We compare our theoretical predictions with recent experiments by Nagy et al.~\cite{nagy2024neutrophils} on neutrophil swelling and migration. In each experiment, several hundred adhered neutrophils were simultaneously and uniformly exposed to chemoattractants. For each cell, velocity was inferred from time-lapse tracking, contact area was measured directly, and volume was quantified using Fluorescence Exclusion Microscopy~\cite{nagy2024neutrophils,Cadart2017}. We begin by summarizing the main experimental findings and interpret them through the lens of our theory. As shown in Fig.~\ref{fig4}, our predictions show excellent agreement with the measured data.

{\bf Cell spreading and swelling -} The cells exhibit two phases in their response to chemoattractant exposure: spreading and swelling (see Fig.~\ref{fig4}). The cell spreads within a couple of minutes, where the contact area increases by about $10\%$, while the volume decreases by around $5\%$. In the following $15$ minutes, the cell swells. The area continues to increase by an additional $20\%$, while the volume increases to approximately $15\%$ more than the initial volume.  The swelling was shown to be associated with ion transport and was suppressed upon inhibition of NHE1 exchangers~\cite{nagy2024neutrophils}, which play an important role in regulatory volume increase ~\cite{Hoffmann2009,Delpire2018,Ni2024}. 

Our theory provides an intuitive explanation of these findings. We model the response to chemoattractants via the two primary drivers of cell dynamics in our theory: (i) a sudden increase in active force, represented by a larger value of $f_0$, and (ii) a rapid change in ion conductances, resulting in an adaptation of the membrane potential, captured by the $\delta_2$ term in Eq.\eqref{psi}. The combination of these two effects drives an increase in the cell area, with the dynamics described by Eq.~\eqref{dAdt}.
Most of the spreading occurs on a timescale of $\sim\tau_{\theta}$, during which the volume decreases due to mechanosensitivity [$\delta_1$ term in Eq.~\eqref{psi}]. Over longer timescales, on the order of $\tau_\psi$, the volume increases as a result of the regulated increase in membrane potential ($\delta_2>0$).

{\bf Cell migration -}
The experiments also reveal an interesting behavior of the cellular velocities. Before the exposure to chemo-attractants, the cells move randomly with a typical velocity of about 2 \textmu m/min \cite{nagy2024neutrophils}. This motion corresponds to a random velocity $u_{\xi}$ that is possible even for $u=0$, due to fluctuations in the active force, as discussed above. Upon exposure, the velocity increases and reaches a new typical value, about twice as large as the original one (see Fig.~\ref{fig4}c), which remains relatively constant on the scale of $10$ minutes. During this time, the cellular motion remains random with low persistence. We regard this as a change in $u_{\xi}$, whose modeling lies beyond the scope of this work.

About $10$ minutes into this new plateau, the cellular velocity increases with time, as the cell continues to swell. Cellular motion also becomes more persistent. We interpret this behavior as the onset of persistent migration with $u>0$, where the total velocity observed in the experiments is $u_{tot}=\sqrt{u^2+u_{\xi}^2}$. The increase in velocity is explained by the increase in contact area during swelling (Fig.~\ref{fig4}a), according to Eq.~\eqref{ueq}. Explicitly, we consider a critical contact area of $A_c\approx1.2 \,A(t=0) $. Once this value is achieved, $u$ smoothly increases with the contact area, explaining the observed behavior. The velocity is monotonically increasing, since the maximal velocity is predicted for $A=9A_c/4$, which was not reached in the experiments. Importantly, these findings highlight that swelling promotes faster migration by increasing the contact area rather than the volume.


\section{Discussion}
This Letter introduces a minimal framework to elucidate cellular size regulation during migration, predicting distinct regimes of migration velocity and its dependence on cell size, including a critical contact area for migration. Our predictions can be tested experimentally by simultaneously tracking cell position, contact area, and volume across various cell types.

Our theory provides an explanation for the increase in migration speed of neutrophils upon swelling, as reported by Nagy et al.~\cite{nagy2024neutrophils}. Swollen cells exhibit larger contact areas, facilitating higher polarization, which in turn increase migration speed and persistence. Crucially, it is the increase in contact area due to the active cellular force that promotes the increase in cell speed, rather than the water content linked to cell volume, as previously suggested. To validate this interpretation, we examined the correlation between neutrophil migration velocity, volume, and contact area (Appendix E). Our findings indicate that migration velocity correlates more strongly with contact area than with volume, supporting our theoretical predictions.

A key distinction arises between the two primary drivers of cell size: membrane potential and active forces. The membrane potential governs intrinsic cell size, inducing simultaneous changes in volume and area. An increase in membrane potential 
increases both cellular volume and area, following the scaling relationship $A\sim V^{2/3}$. In contrast, active forces modulate the contact angle, producing opposing effects on volume and area. An increase in $f_0$  increases the contact area while reducing cellular volume, mediated by ion channels mechanosensitivity. 

This study advances our understanding of the interplay between electrophysiology and mechanics in regulating the concurrent dynamics of cell size and migration. Further experiments across different cell types are required to fully uncover this coupling.


\section{ Acknowledgments.} We thank Tamas L. Nagy, Nir S. Gov, Samuel A. Safran and Kinneret Keren for useful discussions and suggestions.

\appendix
\section{Theory of cellular volume}
Here we derive the volume dynamics (Eqs. (\ref{V(t)},\ref{psi}) in the main text) from the pump-leak model of cellular volume regulation. Cell volume is determined by osmotic pressure balance between the intracellular and extracellular environments, with intracellular pressure regulated via ion transport.

{\bf Pump-leak model.}
The cell is modeled as a solution containing ions that can exchange with the extracellular environment, as well as impermeant molecules such as proteins and amino acids. For simplicity, we consider only the most abundant ions: $\Na^+$, $\K^+$, and $\Cl^-$. On the timescale of tens of minutes, which is the focus of our study, the number of impermeant molecules $N$ is assumed constant. Their average charge is $-ze$, where $-e$ is the electron charge and $z$ is the valency in absolute value. The nucleus is assumed to occupy a fixed fraction of the cell volume~\cite{Romain2022,Deviri2022} and is coarse-grained~\cite{AdarPNAS} in our model.
The charged impermeant molecules, together with active ion pumping (primarily via the Na$^+$/K$^+$ exchanger), establish a membrane potential, $\kbt \psi/e < 0$, which attracts cations and repels anions.

Ion transport between the cell and its surroundings is governed by both passive leak down the electrochemical gradient and active pumping, described by~\cite{Keener,Kay2017} 
\begin{align} 
\label{eqsa1}
\frac{\D}{\D t}\left[n_\intra\left(V-V_d\right)\right]=\frac{A}{z_n e}\left[\frac{\kbt}{e}g_n\left(\frac{1}{z_n}\ln\frac{n_\e}{n_\intra}-\psi\right)+\tilde{p}_n\right]. 
\end{align} 
Here, $n_\intra$ ($n_\e$) is the intracellular (extracellular) molar concentration of ionic species $n$, and $V - V_d$ is the effective cell volume, with $V_d$ denoting the dry (non-aqueous) volume contributed by macromolecules and intracellular membranes (measured experimentally in the limit of infinite extracellular pressure). The ionic valency is $z_n$. Passive transport follows Ohm’s law and is characterized by the electric conductance $g_n$, with the term $\kbt \ln\left(n_\e / n_\intra\right)/ z_n e$ corresponding to the Nernst potential of ion $n$. Active transport is represented by the pumping current density $\tilde{p}_n$. In the standard pump-leak model, only the Na$^+$/K$^+$ exchanger is considered, such that $\tilde{p}_\Na = -3\tilde{p}$, $\tilde{p}_\K = 2\tilde{p}$, and $\tilde{p}_\Cl = 0$.

The (dimensionless) membrane potential enforces electroneutrality and is given by 
\begin{equation} 
\label{eqsa2} 
\psi=\frac{1}{\gt}\left(g_\K\ln\frac{\K_\e}{\K_\intra}+g_\Na\ln\frac{\Na_\e}{\Na_\intra}-g_\Cl\ln\frac{\Cl_\e}{\Cl_\intra}-\tilde{p}\right), 
\end{equation} 
where $\gt = g_\K + g_\Na + g_\Cl$ is the total ionic conductance, and $-\tilde{p}$ is the total electric current driven by the Na$^+$/K$^+$ exchanger.

{\bf Osmotic pressure balance.}
The cell is in mechanical equilibrium with its environment, which is well-approximated by osmotic pressure balance, as the contribution of surface tension is negligible~\cite{Guo2017,AdarPNAS}. The osmotic balance condition (with pressure divided by $\kbt$) reads: 
\begin{equation} 
\label{eqsa3} 
\Na_\intra+\K_\intra+\Cl_\intra+\frac{N}{V-V_d}=\Na_\e+\K_\e+\Cl_\e. 
\end{equation} 
Enforcing electroneutrality inside and outside the cell simplifies this to: 
\begin{equation} 
\label{eqs} 
\left(1+z\right)\frac{N}{V-V_d}+2\Cl_\intra=2\Cl_\e. 
\end{equation}
The left-hand side represents the intracellular osmotic pressure: the first term arises from impermeant molecules and their counterions, and the second term accounts for extraneous salt—$\Cl^-$ ions not required to neutralize the fixed charge—and their counterions.

We obtain a direct relation between cell volume and $\Cl^-$ concentration: 
\begin{align} 
\label{eqsa4} 
v=\frac{V-V_{d}}{V_{m}-V_{d}} &=\frac{1}{1-\Cl_\intra/\Cl_\e}. 
\end{align} 
Here, $V_m = V_d + (1+z)N / 2\Cl_\e$ is the minimal volume in isotonic conditions, corresponding to the limit of an infinitely polarized membrane where all $\Cl^-$ is expelled from the cell.

{\bf Cell volume and membrane potential.}
Because $\Cl^-$ is not actively pumped, its steady-state intracellular concentration is given by $\Cl_\intra = \Cl_\e \exp(\psi)$, leading to 
\begin{equation} 
\label{eqsa5} 
\frac{V-V_d}{V_m-V_d}=\frac{1}{1-e^{\psi}}. 
\end{equation}
We approximate the instantaneous volume using this steady-state expression, which amounts to neglecting the $\Cl^-$ transport timescale~\cite{AdarXiv2024} when $g_\Cl$ is sufficiently large. 

According to Eq.~\eqref{eqsa2}, the electrostatic potential is determined by the ionic conductance values and intracellular concentrations. We assume that the conductance of mechanosensitive (stretch-activated) channels immediately adapts to the instantaneous value of the cellular tension and, similarly, that activation of channels is immediate (step jump at $t=0$). The ionic concentrations, on the other hand, change more slowly and are constrained by typical leak times, as determined by the ionic conductance~\cite{AdarXiv2024}.  We characterize this processes by the potential adaptation time, $\tau_\psi$ ~\cite{AdarXiv2024}. This is the longest timescale of the linearized pump-leak rate equations. 

We approximate the potential dynamics within linear response. Accounting for the changes in conductances and ionic concnentrations, we find that
\begin{equation}
\label{eqs6}
\psi\left(t>0\right)=\psi_0-\delta_1\frac{\gamma-\gamma_0}{\gamma_0}-\delta_2\e^{-t/\tau_\psi},
\end{equation}
where $\psi_0$ is the value of the dimensionless membrane potential at steady state at the end of activation, $\delta_1$ accounts, to linear order, for the effect of mechanosensitivity, where $\gamma$ is the cellular tension and $\gamma_0$ is the reference tension, $\delta_2=\psi_0-\psi(t=0)$ is the potential difference due to activation.

\section{Stability analysis of cellular area and volume}
In this section, we analyze the linear stability of cellular area and volume around steady state, considering small fluctuations in volume and contact angle, $V=V_0+V_1 \exp(st)$ and $\Theta=\Theta_0+\Theta_1(t)\exp(st)$, with $V_1/V_0\ll1$ and $\Theta_1\ll1$ and where we have defined the decay/growth rate $s$. 
Here we do not consider activation ($f_0$ is fixed and $\delta_2=0$). 

Area and volume are coupled by the spherical cap approximation (Eq. (\ref{sphericalcap}) in the main text) and mechano-sensitivity ($\delta_1$ term in Eq. (\ref{psi}) in the main text). Changes in volume and contact angle are induced by changes in cellular surface tension $\gamma=\gamma_0+\gamma_1 \exp(st)$. By linearizing Eqs. (\ref{eq7})-(\ref{dAdt}) in the main text), we find that
 \begin{align}
 \label{eqsb1}
 V_1 &=-\frac{\partial V}{\partial \psi}\bigg|_0\delta_1 \frac{\gamma_1}{\gamma_0},\nonumber\\
 \Theta_1 &=-\frac{1}{1+\tau_{\theta} s}\frac{f_0}{\gamma_0^2} \gamma_1.
\end{align}
 At the same time, the surface tension is determined by the rate of change in total surface area, according to
\begin{align}
\label{eqsb2}
\left(1+\tau_{\gamma}s\right)\gamma_1 &= \eta s\frac{A_{tot,1}(t)}{A_0}\nonumber\\
 &=\eta s \frac{1}{A_0}\left(\frac{\partial A_{tot}}{\partial V}\Big|_0 V_1+\frac{\partial A_{tot}}{\partial \Theta}\Big|_0 \Theta_1\right).
\end{align}
Inserting Eq.~\eqref{eqsb1} in Eq.~\eqref{eqsb2} leads to
\begin{equation}
\label{eqsb3}
\begin{aligned}
\left(1+\tau_{\gamma}s\right)\gamma_1  ={} & -\eta s \frac{1}{A_0} \bigg( 
\frac{\partial A_{tot}}{\partial V}\Big|_0 \frac{\partial V}{\partial \psi}\Big|_0 \delta_1 \\
& + \frac{\partial A_{tot}}{\partial \Theta}\Big|_0 \frac{1}{1+\tau_{\theta}s} \frac{f_0}{\gamma_0} 
\bigg) \frac{\gamma_1}{\gamma_0}.
\end{aligned}
\end{equation}
Note that all the derivatives on the right-hand side are positive. 

The system is linearly unstable if Eq.~\eqref{eqsb3} is solved by $s>0.$ Considering such positive $s>0$ and $\delta_1>0$ as we consider in the main text (membrane potential increases with increasing tension), the prefactor on the right-hand side of Eq.~\eqref{eqsb3} is negative, while that on the left-hand side is positive, leading to a contradiction. This means that $\delta_1>0$ ensures stability.

\section{Analysis of migration velocity}

The force balance equation of a migrating cell is given by the balance of cell-substrate friction and active actin-driven forces,
\begin{equation}
\label{Newton}
m\dot{u}=F_{friction}+F_{actin}
\end{equation}
where $m$ is the cell's mass and $\dot{u}$ is the acceleration.

The friction force is assumed to be a simple drag force proportional to the cell speed,
\begin{equation} \label{F_friction}
F_{friction} = 2\pi R \Gamma u,
\end{equation}
where $\Gamma$ is the cell-substrate friction per unit length.

We consider a net active force along the polarization axis that we set as the $x-$ axis. It is given by
\begin{equation} \label{F_actin}
F_{actin} = R\int_0^{2\pi}f_a(\phi)\,\cos(\phi)\,d\phi
\end{equation}
where $R$ is the radius of the contact area which is related to the contact area by $R=\sqrt{A/\pi}$.

Inserting these forces in Eq.~\eqref{Newton}, we find that 
\begin{equation} \label{Newton2}
 \frac{m}{2\pi\Gamma R}\dot{u}=-u+\frac{1}{2\pi \Gamma}\int_0^{2\pi}f_a(\phi)\,\cos(\phi)\,d\phi .
\end{equation}
At steady state, this reduces to Eq.~\eqref{ForceBalance} of the main text.

We consider a generic feedback between cellular migration and the advection and diffusion of polarity cues within the cytoplasm. 
The advection is assumed to be proportional to the cell's speed ($u\propto v_{advection}$), and the diffusion rate across a unit area within the cell is $D$.
The magnitude of the velocity (or speed) is charactarized by the P\'{e}clet number  $\tilde{u}=uR/D$.

To simplify our treatment of the force distribution, we decompose $f_a(\phi)$ into homogeneous and inhomogeneous components,
\begin{equation} \label{fa_app}
f_a\left(\phi\right) = f_0 \left[1+p(\tilde{u}\,\cos\phi)\right]
\end{equation}
where $f_0$ is an intrinsic scale of force per unit length and $p\left(\tilde{u}\,\cos(\phi)\right)$ accounts for the polarization effect due to the polarity cue asymmetric distribution. As only the latter contributes to a net force, the velocity can be written as
\begin{equation} 
 \frac{m}{2\pi\Gamma R}\dot{u}=-u+\frac{f_0}{2\pi \Gamma}\int_0^{2\pi}p(\tilde{u}\,\cos\phi)\,\cos(\phi)\,d\phi .
\end{equation}

Next, we expand the polarization function $p$ in a Taylor series,  carry out the angular integration, and non-dimensonalize. This leads to
\begin{equation} \label{bifurc}
 \tau_u\dot{\tu}=\frac{f_0 R}{\Gamma D}\left[\left(p_1-\frac{\Gamma D}{f_0 R}\right) \tu+p_3\tu^3+p_5\tu^5+...\right]
\end{equation}
where $\tau_u=m/\left(2\pi\Gamma R\right)$ is a typical time for velocity relaxation and $p_k$ are the expansion coefficients. We consider $\tau_u$ to be smaller than the other timescales of our theory ($\tau_\psi, \tau_\theta,\tau_\gamma$), such that only the steady-state velocity $u$ is considered in the main text. Here, we retain the term $\tau_u\dot{\tu}$ to perform a bifurcation analysis below.

In the following, we will analyze Eq.~\eqref{bifurc} in two cases: (i) $p_3<0$, where an expansion  of order $O\left(\tu^3\right)$ suffices, and (ii) $p_3>0$, where the  $O\left(\tu^5\right)$ order is required. This analysis results in the bifurcation curves presented in Fig.~\ref{fig2} of the main text. 

\textbf{Third order expansion}. Up to the order of $O(u^3)$ the steady-state velocity solves

\begin{equation} \label{u^3}
 \tu=\frac{f_0 R}{\Gamma D}\left(p_1\tu -|p_3|\tu^3\right).
\end{equation}


The non-zero solution of Eq.(\ref{u^3}) in terms of $u$ is given by
\begin{equation}
u=\frac{D}{R}\sqrt{\frac{p_1}{|p_3|}\left(1-\frac{\Gamma D}{p_1 f_0 R}\right)}.
\end{equation}

which we write as
\begin{equation} \label{u0,Rc}
u= u_0\frac{R_c}{R}\sqrt{1-\frac{R_c}{R}},
\end{equation}

whereby the transition from a zero solution ($u=0$) and a non-zero solution ($u\neq0$) occurs at a critical value of 
\begin{equation}
\left(u_c,R_c\right) = \left(0,\frac{\Gamma D}{f_0 p_1}\right),
\end{equation}
and the maximal velocity that the cell can reach occurs at a critical value of
\begin{equation}
\left(u_M,R_M\right) = \left(\frac{2u_0}{3^{3/2}},\frac{3 R_c}{2}\right)
\end{equation}
whereby $u_0=\frac{f_0 p_1}{\Gamma}\left(\frac{p_1}{|p_3|}\right)^{1/2}$.

Overall for the case of $O(\tilde{u}^3)$ we find that 1) for $R<R_c$ the cell is not motile, 2) for $R_c<R<R_M$ the cell is motile with a velocity that increases with the radius, and 3) for $R>R_M$ the cell is motile with a velocity that decreases with the radius. For $R\rightarrow \infty$ the velocity asymptotically reaches zero.
Note that we have assumed for simplicity that $\Gamma$ and $f_0$ are independent of $R$. Our analysis will remain qualitatively the same as long as $\Gamma D/f_0=\theta(R)$.

\textbf{Fifth order expansion}. Up to the order of $O(u^5)$ the velocity can be written as

\begin{equation} \label{u_5}
\tilde{u}=\frac{f_0 R}{\Gamma D}\left(p_1\tilde{u}+p_3\tilde{u}^3-|p_5|\tilde{u}^5\right)
\end{equation}

The non-zero solutions of Eq.(\ref{u_5}) in terms of $u$ are given by

\begin{equation}
u=\frac{D}{R}\sqrt{\frac{p_3}{2|p_5|}\pm\sqrt{\left(\frac{p_3}{2p_5}\right)^2+\frac{p_1}{|p_5|}\left(1-\frac{R_c}{R}\right)}}
\end{equation}

which can be written as
\begin{equation} \label{eq:uu5}
    u=u_c^*\frac{R_c^*}{R}\sqrt{1\pm\sqrt{\frac{R_c}{R_c-R_c^*}\left(1-\frac{R_c^*}{R}\right)}}
\end{equation}

whereby the non-smooth transition (first order phase transition) from a zero solution ($u=0$) and a non-zero solution ($u\neq0$) occurs at a critical value of

\begin{equation}
\begin{aligned}
R_c^* &= R_c/\left(1+\frac{p_3^2}{4p_1|p_5|}\right) \\
u_c^* &= u_0\frac{R_c}{R_c^*} \sqrt{\frac{p_3^2}{2p_1|p_5|}}=u_0\frac{R_c}{R_c^*} \sqrt{2\left(1-\frac{R_c}{R_c^*}\right)} 
\end{aligned}
\label{eq:Rcuc}
\end{equation}

Overall for the case of $O(\tilde{u}^5)$ we find that 1) for $R<R_c^*$ the cell is not motile, 2) for $R_c^*<R<R_c$ the cell exhibit a bi-stable behavior of $u=0$ and a non-zero $u$ which increases with the cell size. 3) For $R_c<R<R_M^*$ the cell is motile with a velocity that increases with size, and 4) for $R>R_M^*$ the cell is motile with a velocity that decreases with the radius. For $R\rightarrow \infty$ the velocity asymptotically reaches zero.


\section{Analysis of experimental results: cell velocity depends on area rather than volume}
In the appendix we present the correlations between the contact area, volume, and velocity data, measured by Nagy et al. \cite{nagy2024neutrophils}, and available via DataDryad (see https://doi.org/10.7272/Q6NS0S5N.)

For this data we calculate the pairwise correlation coefficients, defined by
\begin{equation}
r(x,y) = \frac{\sum_{i=1}^{n}(x_i - \bar{x})(y_i - \bar{y})}
{\sqrt{\sum_{i=1}^{n}(x_i - \bar{x})^2}\sqrt{\sum_{i=1}^{n}(y_i - \bar{y})^2}},
\end{equation}
where $\bar{x}$ and $\bar{y}$ are the sample means of the vectors $x$ and $y$, respectively, and $n$ is the number of observations.

\begin{figure}[ht]
\centering
\includegraphics[width=1\columnwidth]{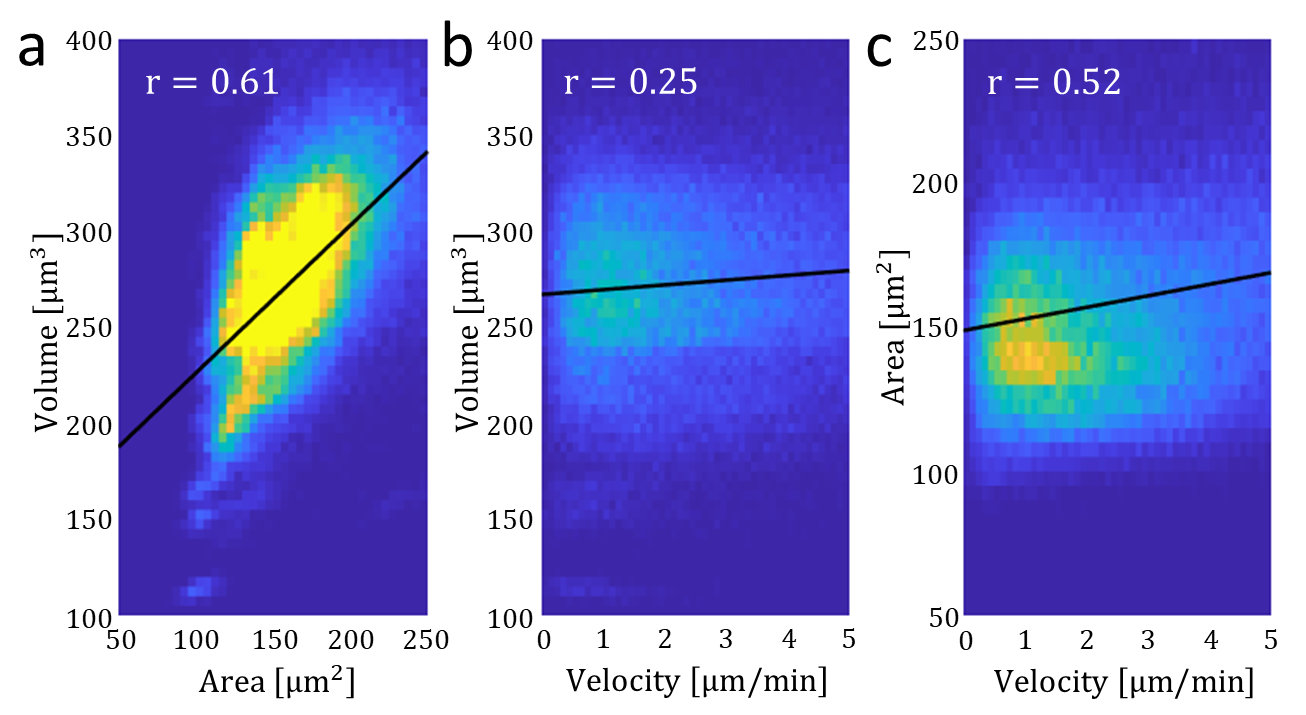}
\caption{Correlation between Volume speed and area (data taken from Nagy et al. \cite{nagy2024neutrophils}). A) Volume and Area. B) Volume and velocity. C) Area and velocity. Black line is the linear fit. $r$ denotes the pairwise correlation coefficients.}
\label{fig5}
\end{figure}

As expected we find a strong correlation between the volume and the area (Fig.\ref{fig5}a). We also find that the correlation between the velocity and the volume (Fig.\ref{fig5}b) is significantly smaller than the correlation between the velocity and the area (Fig.\ref{fig5}c), which aligns with our theory. 
%


\end{document}